\pdfoutput=1
\documentclass[11pt]{article}

\usepackage{graphicx}
\usepackage{amsmath,amssymb}
\usepackage{dcolumn}
\usepackage{bm}
\usepackage{slashed}
\usepackage[bookmarks,bookmarksnumbered,colorlinks=true,anchorcolor=blue,
linkcolor=blue,urlcolor=blue,citecolor=blue,breaklinks=true]{hyperref}
\usepackage[utf8]{inputenc}
\usepackage{authblk}

\usepackage{color}
\usepackage{ulem}

\newcommand{\be}{\begin{equation}}
\newcommand{\ee}{\end{equation}}
\newcommand{\ba}{\begin{eqnarray}}
\newcommand{\ea}{\end{eqnarray}}
\def\bea{\begin{eqnarray}}
\def\eea{\end{eqnarray}}

\newcommand{\gsim}{\mathrel{\hbox{\rlap{\lower.55ex \hbox {$\sim$}}
                   \kern-.3em \raise.4ex \hbox{$>$}}}}
\newcommand{\lsim}{\mathrel{\hbox{\rlap{\lower.55ex \hbox {$\sim$}}
                   \kern-.3em \raise.4ex \hbox{$<$}}}}

\def\roughly#1{\mathrel{\raise.3ex\hbox{$#1$\kern-.75em%
\lower1ex\hbox{$\sim$}}}}
\def\lsim{\roughly<}
\def\gsim{\roughly>}

\def\({\left(}
\def\){\right)}
\def\[{\left[}
\def\]{\right]}
\def\<{\langle}
\def\>{\rangle}

\def\a{{\alpha}}
\def\b{{\beta}}

\def\m{{\mu}}
\def\n{{\nu}}

\usepackage{xcolor}

\title{\bf {Thermodynamical property of entanglement entropy and deconfinement phase transition}}

\author[1]{Mitsutoshi Fujita\thanks{fujita@mail.sysu.edu.cn}}
\author[2,3]{Song He\thanks{hesong@jlu.edu.cn }}
\author[1]{Yuan Sun\thanks{sunyuan6@mail.sysu.edu.cn}}
\affil[1]{School of Physics and Astronomy, Sun Yat-Sen University, Guangzhou 519082, China}
\affil[2]{Center for Theoretical Physics and College of Physics, Jilin University, Changchun 130012, People’s Republic of China}
\affil[3]{Max Planck Institute for Gravitational Physics (Albert Einstein Institute), Am M\"uhlenberg 1, 14476 Golm, Germany}
\renewcommand\Authands{ and }

\date{}

\begin{document}

\maketitle

\begin{abstract}

\end{abstract}
{We analyze the holographic entanglement entropy in a soliton background with Wilson lines and derive a relation analogous to the first law of thermodynamics. The confinement/deconfinement phase transition occurs due to the competition of two minimal surfaces. The entropic c function probes the confinement/deconfinement phase transition.
It is sensitive to the degrees of freedom (DOF) smaller than the size of a spatial circle. When the Wilson line becomes large, the entropic c function becomes non-monotonic as a function of the size and does not satisfy the usual c-theorem. We analyze the entanglement entropy for a small subregion and the relation analogous to the first law of thermodynamics. For the small amount of Wilson lines, the excited amount of the entanglement entropy decreases from the ground state. It reflects that confinement decreases degrees of freedom. We finally discuss the second order correction of the holographic entanglement entropy. }

\newpage

\allowdisplaybreaks

\flushbottom

\section{Introduction}
Entanglement entropy of a subsystem $A$ counts the number of degrees of freedom of the quantum entangled state in quantum field theories~\cite{Cardy,Review}.
In the condensed matter physics, it is divergent at the critical point for quantum critical phase transitions and becomes an order parameter~\cite{Vidal:2002rm}. It captures  geometric discernment of field theories such as an area law~\cite{EE}: the entanglement entropy defined in a subregion looks like the black hole entropy.

The Ryu-Takayanagi formula proposes the holographic dual of the entanglement entropy~\cite{Ryu:2006bv,Ryu:2006ef,Nishioka:2009un}. It is a powerful tool to analyze strongly coupled systems. It has been the order parameter of the confinement/deconfinement phase transition in a confining gauge theory~\cite{Nishioka:2006gr,Klebanov:2007ws,Buividovich:2008gq,Dudal:2016joz,Dudal:2018ztm,Mahapatra:2019uql}. The phase transition occurs due to the competition of two minimal surfaces. After the phase transition, the entanglement entropy turns out to be trivial in the confined phase at the infrared red limit. The holographic entanglement entropy (HEE) also probes holographic superconductor phase transitions~\cite{Albash:2012pd}-\cite{Das:2017gjy}.

The entanglement entropy for excited states has been attracting attentions. The entanglement entropy in the small region satisfies a relation similar to the first law of thermodynamics~\cite{Bhattacharya:2012mi}
\ba
T_{ent}\Delta S_A =\Delta E_A,
\ea
where $\Delta E_A$ is the increased amount of energy in the subregion and $\Delta S_A$ is the increased amount in excited states compared with the ground state of a CFT. $T_{ent}$ is called the entanglement temperature. This relation has been investigated in many holographic models dual to field theories at finite temperature~\cite{Guo:2013aca,Allahbakhshi:2013rda}. There are very extensive investigations \cite{He:2013rsa}\cite{Park:2015hcz}\cite{Caceres:2016xjz}\cite{Lin:2016fua}\cite{Ghosh:2016fop}\cite{Sun:2016dch}\cite{OBannon:2016exv}\cite{Bhattacharya:2017gzt}\cite{Bhattacharya:2019zkb}\cite{Lokhande:2017jik} on the first law like relation of the holographic entanglement entropy in various cases.

The second order correction to the holographic entanglement entropy has been studied by \cite{Blanco:2013joa}. In the paper, authors rewrite the first law like relation of entanglement entropy in terms of the Relative entropy \cite{Blanco:2013joa}. They calculated the first law like relation of entanglement entropy with spherical entanglement surface up to second order and they also took the deformation of entanglement surface into account. For spherical entanglement surface, authors extend the first law relation to higher order and give insights to use boundary information \cite{Lin:2014hva} to reconstruct the bulk geometry. However, it is very difficulty to study the second order correction to the holographic entanglement entropy for generic entanglement surface, for example, strip case \cite{He:2014lfa} and so on. In Ref. \cite{He:2014lfa}, author have studied the strip entanglement surface up to the second order corrections from gravitational background without taking the shape deformation of the entanglement surface into account. For higher derivative gravity, the dictionary of the holographic entanglement entropy has to be changed in terms of \cite{Dong:2013qoa}. In higher derivative gravity theories, the second order of holographic entanglement entropy becomes very complicated. In the literature, authors \cite{Pal:2015mda}\cite{Sun:2016til}\cite{Bueno:2016gnv}\cite{Haehl:2017sot} have studied the similar first law like relation of the entanglement entropy in various situations. Further, the corresponding holographic entanglement chemistry is investigated by \cite{Caceres:2016xjz}\cite{Gushterov:2017vnr}. It would be interesting to apply to confining gauge theories with Wilson lines. In an $AdS$ soliton background, the boundary energy becomes negative and comparable with the negative Casimir energy of a confining gauge theory. Adding Wilson lines will vary the boundary energy.

In this paper, we analyze a phase transition as well as thermodynamic properties of the holographic entanglement entropy in a solitonic background with the current and the Wilson line. We will show that the Wilson line increases the energy and becomes even positive. The vacuum expectation value of the current also excites the state. We consider both small and large subregions of the entanglement entropy. We compute the entanglement entropy and the entropic c-function. The latter becomes a nice probe of the confinement/deconfinement phase transition. We will demonstrate that a phase transition occurs for the Kaluza-Klein mass(mass of a Kaluza-Klein state)$\sim 1/l$ analogous to~\cite{Nishioka:2006gr}. We also analyze the contribution of the Wilson line and the current to the entropic c-function \cite{Nishioka:2006gr}. We analyze the entanglement entropy at the small subregion. Computing the boundary energy and increase amount of the entanglement entropy, we will obtain an entanglement temperature. As a byproduct, we work out the generic formula for the second order correction to the holographic entanglement entropy with contributions from the deformation of the entangling surface. As a consistency check, we apply this formula to a spherical entangling surface and the resulting second order corrections are the same as ones presented in \cite{Blanco:2013joa}. One can do the similar investigation in the strip case with some numerical simulation. We leave the problem in the future work.

In section \ref{sec1}, we compute free energy of a QFT dual to a solitonic background with the current. In section \ref{sec2}, we compute quasi-local stress tensor of the solitonic background. In section \ref{sec3}, we compute the holographic entanglement entropy with a striped region. We analyze the confinement/deconfinement phase transition. We also introduce the entropic c-function to probe a phase transition.  For the small subregion, we compute the relation as in the first law of the thermodynamics.
 In section \ref{2NDO}, we compute the second order correction to the holographic entanglement entropy with spherical entanglement surface. In appendices, we would like to list some relevant techniques and notations which are very useful in our analysis.

\section{Free energy}\label{sec1}
In this section, we compute free energy of a QFT with Wilson lines by using the gauge/gravity correspondence.
The gravitational action with the Maxwell field has $U(1)$ gauge symmetry, which corresponds to $U(1)$ global symmetry in the field theory side. The action also includes the Gibbons-Howing boundary term to have a correct variation principle as follows~\cite{Hartnoll:2009sz}:
\ba
&S=\int d^{d+1}x\sqrt{-g}\Big(\dfrac{1}{2\kappa^2}\Big(R+\dfrac{d(d-1)}{L^2}\Big)-\dfrac{1}{4g_e^2}F^2\Big) \nonumber \\
&+\dfrac{1}{2\kappa^2}\int d^dx\sqrt{\gamma}\Big(2K-\dfrac{2(d-1)}{L}\Big),
\ea
where $\gamma$ is the induced metric at the $AdS$ boundary, $K=\gamma^{\m\n}\nabla_\m n_\n$, and a boundary term is added.

The Einstein equations of motion become
\be
R_{\m\n}-\frac{R}{2}g_{\m\n}-\frac{d(d-1)}{2L^2}g_{\m\n}=\frac{\kappa^2}{2g_e^2}(2F_{\m\sigma}F_\n^{~\sigma}-\frac{1}{2}g_{\m\n}F^2).
\ee
We also have the Maxwell equation.



The metric of a $AdS_{d+1}$ soliton becomes a solution of EOM as follows:
\be \label{Sol2}
ds^2_{d+1}=\frac{L^2}{z^2}\Big(-dt^2+\frac{dz^2}{f_d(z)}+f_d(z) d\phi^2+\sum^{d-2}dx^i dx^i\Big),
\ee
where
\be
f_d(z)=1-\Big(1+\frac{\epsilon_1 z^2_+a_\phi^2}{\gamma^2}\Big)\Big(\frac{z}{z_+}\Big)^d+\frac{\epsilon_1 z^2_+a_\phi^2}{\gamma^2}\Big(\frac{z}{z_+}\Big)^{2(d-1)},
\ee
where $a_\phi$ describes Wilson lines and $\epsilon_1=-1$. Recall that $\epsilon_1 =1$ in the Reissner Nordstrom $AdS$ black hole~\cite{Hartnoll:2009sz}. We have described $\gamma^2=\frac{(d-1)g_e^2 L^2}{(d-2)\kappa^2}$, which is a dimensionless combination.
The background gauge field becomes
\be
A_\phi=a_\phi \Big(1-\Big(\frac{z}{z_+}\Big)^{d-2}\Big),
\ee
where $z_+$ is regular at the tip of the soliton. The dual current is defined as $J^\phi =\frac{\delta S}{\delta a_\phi}$.
The signature of $t$ does not affect the background gauge field unlike the Reissner Nordstr{\"o}m $AdS$ black hole.  Kaluza-Klein mass of the $\phi$ circle is obtained in an Euclidean signature solution as follows:
\ba
M_0=\dfrac{1}{4\pi z_+}\Big(d-\dfrac{\epsilon_1 (d-2)z_+^2a_\phi^2}{\gamma^2} \Big)>0.
\ea
Because a Wilson loop $\oint A$ vanishes around the vanishing circle at $z=z_+$, the gauge connection is regular at the tip of the soliton. Note that $M_0$ becomes non-zero for any real $a_\phi$ and the dimensionless ratio $a_\phi/M_0$ can smoothly be taken to be zero.

We have two branches solving the above equation in terms of $z_+$ as follows:
\ba\label{RPL113}
z_+=\dfrac{-2\pi M_0\gamma^2\pm \sqrt{(2\pi M_0\gamma^2)^2+d(d-2)\epsilon_1 \gamma^2a_\phi^2}}{(d-2)\epsilon_1 a_\phi^2}.
\ea
The solution exists when $|a_\phi|\le \frac{2\pi M_0\gamma }{\sqrt{d(d-2)}}$. Choosing the minus sign in the above formula, $z_+$ is divergent at small $a_\phi$ limit. Since this background does not smoothly continue to the $AdS_{d+1}$ soliton, it is not relevant for our analysis.

 The free energy of the dual field theory is derived from analyzing the Euclidean action via an analytic continuation $Z=e^{-\beta F}=e^{-S_E(g_*)}$. The free energy becomes
\be
F=-\frac{L^{d-1}}{2\kappa^2 z_+^d}\Big(1+\epsilon_1\frac{z^2_+a_\phi^2}{\gamma^2}\Big)V_{d-1}.
\ee
Note that $L^{d-1}/\kappa^2$ is a dimensionless parameter and scales as in a power of $N$.

One can show that the solution of the plus sign is always stabler than that of the minus sign ($d\le 9$). We define a new parameter $a_\phi = \frac{2\pi M_0\gamma }{\sqrt{d(d-2)}}x$ ($|x|\le 1$). In $d=4$ and $5$, especially, the difference is computed as
\ba
\dfrac{\kappa^2}{L^{d-1}M_0^d}(F_--F_+)=\begin{cases} &\dfrac{64}{27} \pi ^3 \Big(1-x^2\Big)^\frac{3}{2}\ge 0\quad (d=3),  \\
& \pi ^4 \Big(1-x^2\Big)^{\frac{3}{2}}\ge 0\quad (d=4).
 \end{cases}
\ea
Thus, we choose the plus sign in \eqref{RPL113} in later study.

\section{Boundary stress tensor}\label{sec2}
In this section we compute the stress tensor of boundary field theory dual to soliton background in two different ways. In the first method the Brown-York tensor with counter terms is used, and in the second the stress tensor is read from FG expansion of metric near the boundary.
Let us begin with the first way. The soliton metric (\ref{Sol2}) with  can be casted into the form (contemplate $d=4$ here)
\be
ds^2=F(z)dz^2+h_{ab}(z)dx^adx^b,~~x^a=t,x^i
\ee
with
\be
h_{ij}=\dfrac{L^2}{z^2}\mbox{diag}(-1,f(z),1,1).
\ee
The boundary stress tensor near the boundary denoted by $\partial M$ (constant-$z$ surface with $z\to 0$) is \cite{Balasubramanian:1999re}
\be\label{ENE18}
T_{ij}=\dfrac{1}{\kappa^2}\Big(K_{ij}-Kh_{ij}-\dfrac{3}{L}h_{ij}\Big)\Big|_{z\to 0},
\ee
where $K_{ij}$ is the extrinsic curvature of the boundary $\partial M$. The first two terms are Brown-York tensor terms, and the last term is a counter term added to yield finite answer near the boundary. Substituting into the metric (\ref{Sol2}), the stress tensor can be derived. Let us focus on the $tt$-component which is relevant in the computation of boundary energy
\be
T_{tt}=-\frac{L\bar{a}_\phi}{2\kappa^2 z_+^4}z^2+O(z^3),~~
 \bar{a}_\phi =1-\Big(\frac{z_+ a_\phi}{\gamma}\Big)^2.
\ee
 It follows that the boundary energy is then (eq.(12) in \cite{Balasubramanian:1999re})
\be
M=\int d^2xd\phi \sqrt{\sigma} u^a \xi^b T_{ab}=\int d^2xd\phi \sqrt{\sigma} u^t \xi^t T_{tt}=-\frac{V_2}{M_0}\frac{L^3\bar{a}_\phi}{2\kappa^2  z_+^4}
\ee
where  $\sigma_{ij}$ is the metric of a spacelike surface $\Sigma$ in $\partial M$, $u^\mu$ is the timelike unit vector normal to $\Sigma$.  $\xi^\mu$ is timelike Killing vector generating time translation isometry of the boundary.
Here $\sqrt{\sigma}=\frac{\sqrt{f}L^3}{z^3}$, $u^t=\frac{z}{L},\xi^t=1,\int d^2x=\int dx^1dx^2=V_2.$ Note that the energy is negative when $a_\phi < a_0=\frac{2\pi M_0\gamma}{d-1}$. When $a_\phi =0$, the above negative energy was compared with the negative Casimir energy of the gauge theory on $S^1\times R^2$. The result has a good agreement with\cite{Horowitz:1998ha}.
 For metric (\ref{Sol2}) in general dimensions, we have
\be
h_{ij}=\dfrac{L^2}{z^2}\mbox{diag} (-1,f(z),1,\dots ,1).
\ee
The $tt$-component of the quasi-local stress tensor becomes
\ba
T_{tt}=-\dfrac{L\bar{a}_\phi}{2\kappa^2 z_h^d}z^{d-2}+O(z^{d-1}).
\ea
The boundary energy is computed as follows:
\ba\label{Ttt1}
M=\int d^{d-2}xd\phi \sqrt{\sigma }u^t\xi^tT_{tt}=-\dfrac{V_{d-2}}{M_0}\frac{L^{d-1}\bar{a}_\phi}{2\kappa^2 z_+^d },
\ea
where $\sqrt{\sigma}=\frac{\sqrt{f}L^{d-1}}{z^{d-1}}$, $\int d^2x=V_{d-2}.$ The energy in general dimensions is also negative when $a_\phi <a_0=\frac{2\pi M_0\gamma}{d-1}$. When $a_\phi=a_0$, the energy vanishes.

Alternatively, we can obtain boundary stress tensor by resorting to FG expansion~\cite{Henningson:1998gx,de Haro:2000xn}. The bulk metric in FG gauge can be written as
\be\label{FGexp}
ds^2=\frac{L^2}{\tilde{z}^2}(d\tilde{z}^2+g_{\m\n}dx^\m dx^\n)
,~~
g_{\m\n}=\eta_{\m\n}+\delta g_{\m\n}
.\ee Here $\eta_{\m\n}$ is Minkowski flat metric,
$\delta g_{\m\n}$ begins with terms of order $\tilde{z}^d$ near the boundary.

Next transforming the soliton background (\ref{Sol2}) into the form (\ref{FGexp})
\be\label{gttFG}
ds^2=\frac{L^2}{z^2}(-dt^2+\frac{dz^2}{h(z)}+h(z)d\phi^2+\sum^{d-2}dx^i dx^i)=\frac{L^2}{\tilde{z}^2}\Big(d\tilde{z}^2+g_{\m\n}dx^\m dx^\n\Big)
\ee
with asymptotic expansion
\be
h=1-\a z^d+\b z^{2d-2},~~\a=\frac{1}{z^d_+}(1-\frac{z^2_+ a^2_\phi}{\gamma^2}),~~\beta=-\frac{1}{z_+^{2d-4}}\frac{a_\phi^2}{\gamma^2}.
\ee
we obtain
\be
g_{tt}=-\Big( 1+\a\frac{\tilde{z}^d}{d}+\b\frac{\tilde{z}^{2d-2}}{2(1-d)}\Big),
\ee
\be
g_{\phi\phi}= 1+\b\frac{(2d-3)\tilde{z}^{2d-2}}{2(d-1)}+\a(-1+\frac{1}{d})\tilde{z}^d,
\ee
\be
g_{ij}=\frac{\tilde{z}^2}{z^2}=\delta_{ij}\Big(1+\a\frac{\tilde{z}^d}{d}+\b\frac{\tilde{z}^{2d-2}}{2(1-d)}\Big).
\ee
The general variation of metric  ,to leading order, takes the form  \cite{Blanco:2013joa}
\be\label{delgGF}
\delta g_{\m\n}=a \tilde{z}^d T^{(0)}_{\m\n}+\tilde{z}^{2d-2}(bJ_\m J_\n+c\eta_{\m\n}J^2),
\ee
where $T^{(0)}$ is the boundary stress tensor, and the boundary current $J_\m$ appears due to the dual gauge field of the bulk (\ref{Sol2}) is turned on.
Here $a=\frac{2\kappa^2}{dL^{d-1}}$ as given in \cite{Blanco:2013joa}, then from (\ref{delgGF}), we can read off all components of stress tensor
\be \label{Ttt2}
T^{(0)}_{tt}=-T^{(0)}_{x^ix^i}= -\frac{\a}{a d}=-\frac{L^{d-1}}{2\kappa^2}\frac{1}{z^d_+}\Big(1-\frac{z^2_+ a^2_\phi}{\gamma^2}\Big)=-\frac{L^{d-1}}{2\kappa^2}\frac{1}{z^d_+}\bar{\a}_\phi,
\ee

\be
T^{(0)}_{\phi\phi}=\frac{\a}{a}\Big(-1+\frac{1}{d}\Big)=\frac{dL^{d-1}}{2\kappa^2}\frac{1}{z^d_+}\Big(1-\frac{z^2_+ a^2_\phi}{\gamma^2}\Big)\Big(-1+\frac{1}{d}\Big).
\ee
Note the (\ref{Ttt2}) is energy density of boundary field theory which is consistent with (\ref{Ttt1}).
\section{The holographic entanglement entropy}\label{sec3}
We compute the holographic entanglement entropy~\cite{Ryu:2006bv,Ryu:2006ef} in this background. We divide the boundary region into two regions. The first region is defined as $-l/2\le x^1\le l/2$ and $0\le x^i\le L_x$ for the remaining $x^i$, and wrapping $\phi$ direction. The second region is the complement. The boundary of the Ryu-Takayanagi surface ends on the boundary of the above region. The surface becomes a codimension 2 surface at a constant time slice with the embedding scalar $z=z(x^1)$. The surface action becomes
\ba\label{ACT111}
A=L_x^{d-2}\int dx^1\Big(\dfrac{L}{z}\Big)^{d-1}\sqrt{(z')^2+f_d(z)}.
\ea
The Hamiltonian of $A$ becomes a constant independent of $x_1$. It leads to the following EOM of the first order
\ba\label{zst112}
z'=\sqrt{f_d(z)\Big(\dfrac{f_d(z)}{f_d(z_*)}\dfrac{z_*^{2(d-1)}}{z^{2(d-1)}}-1\Big)},
\ea
where $z_*$ is the turning point. $z'=0$ at $z=z_*$.
By integrating $z'$, we require the boundary condition
\ba
l=2\int^{z_*}_{\epsilon}dz\dfrac{1}{\sqrt{f_d\Big(\dfrac{f_d(z) z_*^{2(d-1)}}{f_d(z_*)z^{2(d-1)}}-1\Big)}}.
\ea
The above formula relates $l$ with $z_*$.
Substituting \eqref{zst112}, the surface action \eqref{ACT111} becomes
\ba\label{ACT27}
A=2L_x^{d-2}\int^{z_*}_{\epsilon}dz\dfrac{L^{d-1}z^{2-2d}}{\sqrt{z^{2-2d}-\dfrac{f_d(z_*)}{f_d(z)}z_*^{2-2d}}},
\ea
where $\epsilon$ is a small cutoff scale.
The singular part of $A$ becomes $A\sim \dfrac{2L_x^{d-2}L^{d-1}}{(d-2)\epsilon^{d-2}}$.

For pure $AdS$, the surface action \eqref{ACT27} can be integrated over a region. Replacing $z_*$ with $l$, it becomes
\ba\label{ACT28}
A=2L_x^{d-2}L^{d-1}\Big(\dfrac{1}{(d-2)\epsilon^{d-2}}-\dfrac{2^{d-2}\pi^{\frac{d-1}{2}}}{d-2}\Big(\dfrac{\Gamma (\frac{d}{2(d-1)})}{\Gamma (\frac{1}{2(d-1)})}\Big)^{d-1}\dfrac{1}{l^{d-2}} \Big).
\ea

\subsection{The confinement/deconfinement transition}
According to~\cite{Nishioka:2006gr,Klebanov:2007ws}, the holographic entanglement entropy can capture the confinement/
deconfinement phase transition without black brane solutions. In this section, we analyze the confinement/deconfinement transition applying it. We also examine the dependence on the Wilson line along $\phi$. The entanglement entropy counts the degrees of freedom of the entangled states at the energy scale $\Lambda \sim 1/l$. In confining gauge theories, the  behavior of the entanglement entropy will become trivial when $l$ becomes large. That is, it corresponds to the IR limit.

We have two choices of the minimal surface. The first one is a connected surface \eqref{ACT27}, which corresponds to the deconfinement phase. The second one is a disconnected surface, which goes straight from the $AdS$-soliton boundary to the bulk. Because the disconnected surface does not depend on $l$, it corresponds to the confinement phase.

\begin{figure}[htbp]
     \begin{center}
          \includegraphics[height=5cm,clip]{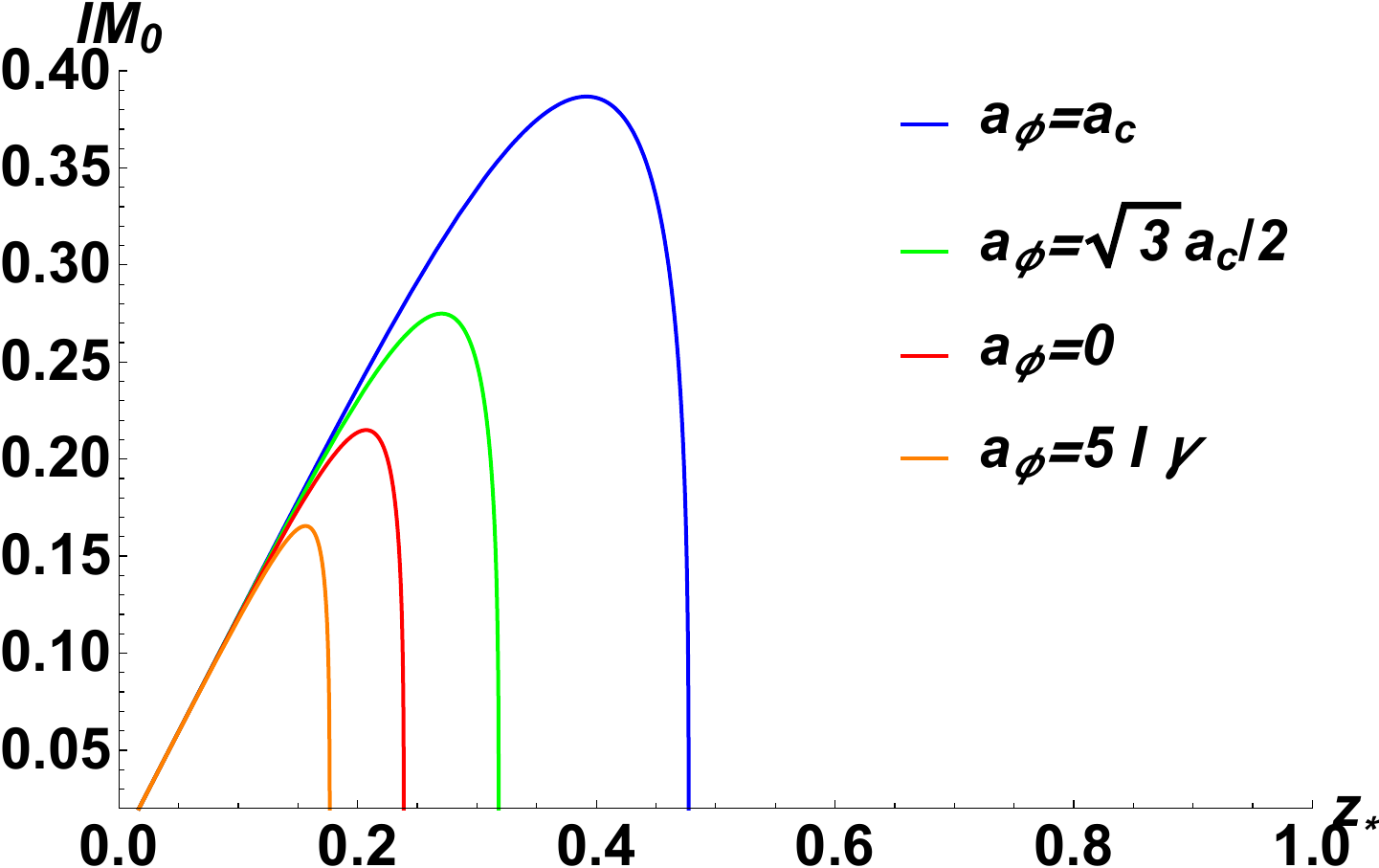}
           \caption{The size $lM_0$ of the subregion with some fixed $a_\phi$ as a function of $z_*$ in $d=3$. At the special value $a_\phi=a_0$, the boundary energy is zero. In the figure $a_0=\sqrt{3}a_c/2$ ($a_c=2\pi M_0\gamma /\sqrt{d (d-2)}$).  The size $l$ linearly grows for small $z_*$ and has the maximal value $l_{max}=0.39,\ 0.27,\ 0.22,\ 0.17$ in units of $1/M_0$ from the top to the bottom, respectively. The figure shows that $z_*$ has two values.  }
    \label{fig:EE}
    \end{center}

\end{figure}
\begin{figure}[htbp]
     \begin{center}
          \includegraphics[height=5cm,clip]{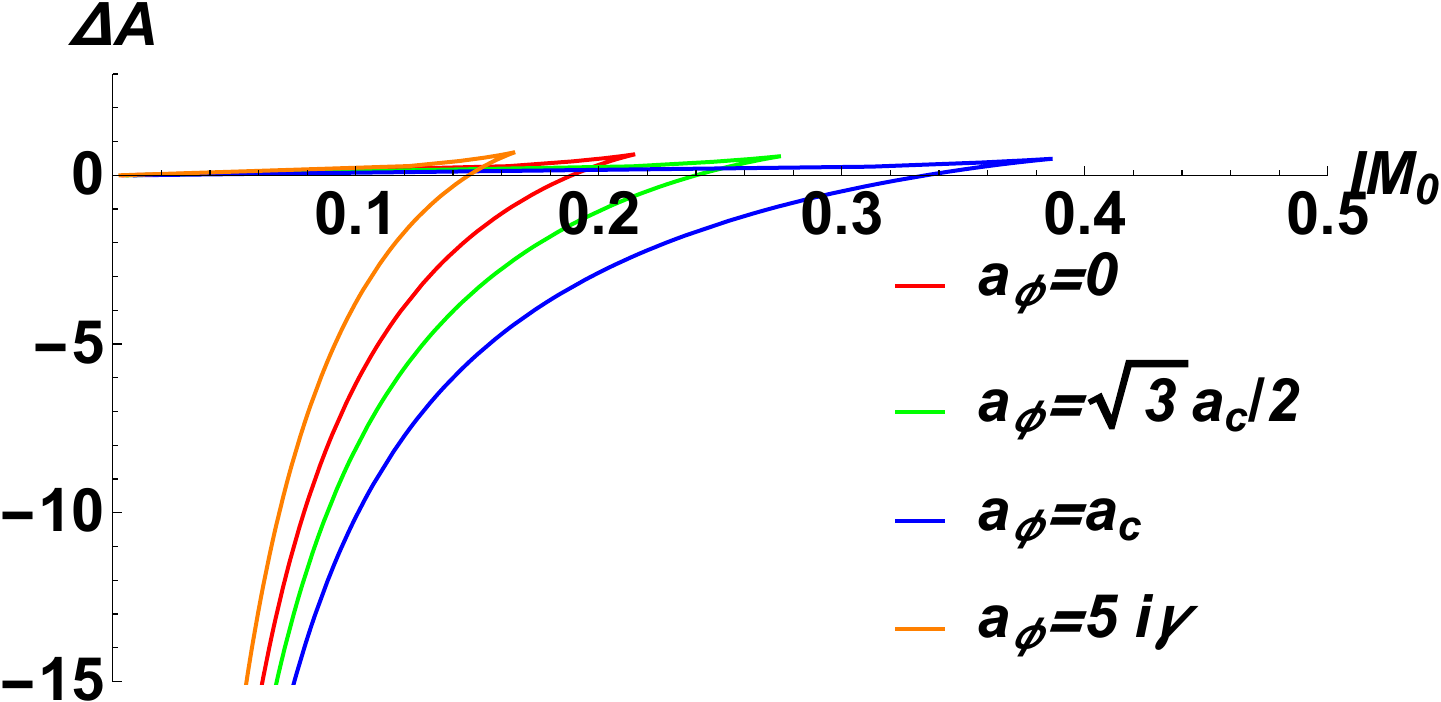}
           \caption{The entanglement entropy as a function of $l$ in 3 dimensions.
The larger surface is unphysical between two connected surfaces of the same curve. When $0<l<l_c$, the connected surface is favored. When $l>l_c$, the disconnected surface dominates the behavior. The critical length is given by $l_c=0.34,\ 0.24,\ 0.19,\ 0.15$ in units of $1/M_0$ from the bottom to the top, respectively.  This implies that the phase transition occurs at the large Kaluza-Klein mass $M_0$ with increase of the Wilson line $a_\phi$.}
    \label{fig:DA}
    \end{center}
\end{figure}

\begin{figure}[htbp]
     \begin{center}
   \includegraphics[height=5cm,clip]{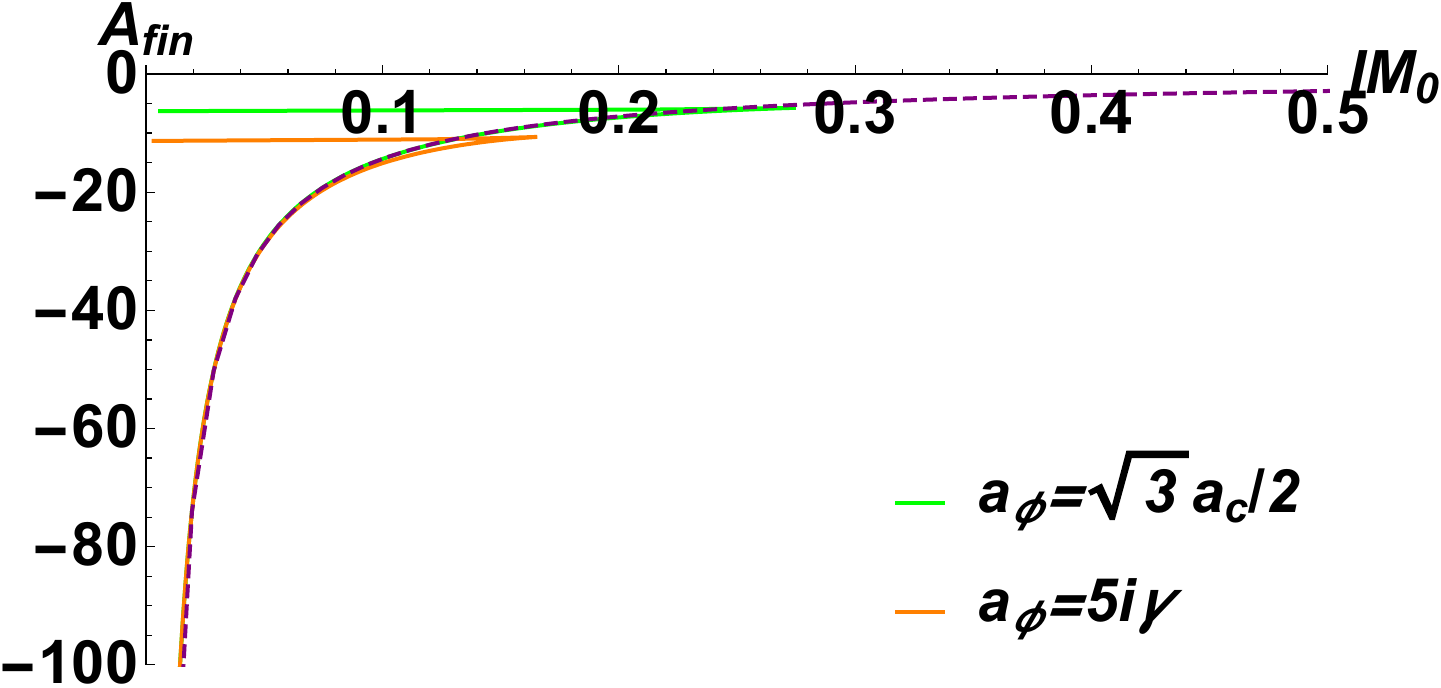}
           \caption{The finite part of the entanglement entropy is plotted as a function of $lM_0$ in 3 dimensions. The curve of $a_\phi =a_c$ almost agrees with that of the pure $AdS$ \eqref{ACT28} when $0<l<l_c$. }
    \label{fig:Afin}
    \end{center}
\end{figure}

The connected surface \eqref{ACT27} has the maximal size $l_{max}$ of the interval, which depends on the Wilson line $a_\phi$ and the Kaluza-Klein mass $M_0$. The size $lM_0$ of the interval is plotted as a function of $z_*$ in 3 dimensions in Fig. \ref{fig:EE}. The size $lM_0$ has the maximal value $l_{max}M_0\sim 1$. Explicitly, $l_{max}M_0=0.39$ when $a_\phi=a_c$. This is larger than that of the $AdS_5$ soliton. When $a_\phi$ becomes imaginary, the curve $a_\phi =5 i\gamma$ leads to the result of the geometric entropy~\cite{Allahbakhshi:2013wk}. The geometric entropy is related to the entanglement entropy via the double Wick rotation~\cite{Fujita:2008zv,Bah:2008cj}.

When $l$ becomes large, the connected surface doesn't exist anymore. Instead, the disconnected surface dominates the behavior. It ends at the tip of the soliton ($z=z_+$).
\ba
A^{dis}=2L_x^{d-2}L^{d-1} \int^{z_+}_{\epsilon} dz \dfrac{1}{z^{d-1}}=\dfrac{2L_x^{d-2}L^{d-1}}{d-2} \Big(\dfrac{1}{\epsilon^{d-2}}-\dfrac{1}{z_+^{d-2}}  \Big).
\ea

The difference $\Delta A=A-A^{dis}$ in 3 dimensions is plotted in Fig. \ref{fig:DA}. There are two connected surfaces of the same curve. The larger one is unphysical. For large $l$, the disconnected surface dominates the behavior. There is a first order phase transition at a critical point $l=l_c$. The critical length increases with increase of $a_\phi$ in general.

To probe  the confinement/deconfinement phase transition, we introduce the entropic c function. It was proposed in~\cite{Nishioka:2006gr}. The entropic c function is defined as
\ba\label{ENT41}
C(l)=\dfrac{l^{d-1}}{V}\dfrac{dS}{dl},
\ea
where $V=L_x^{d-2}$. This is the generalization of the 2 dimensional entropic c function defined in~\cite{Casini:2004bw,Casini:2006es}.
\begin{figure}[htbp]
     \begin{center}
  \includegraphics[height=6cm,clip]{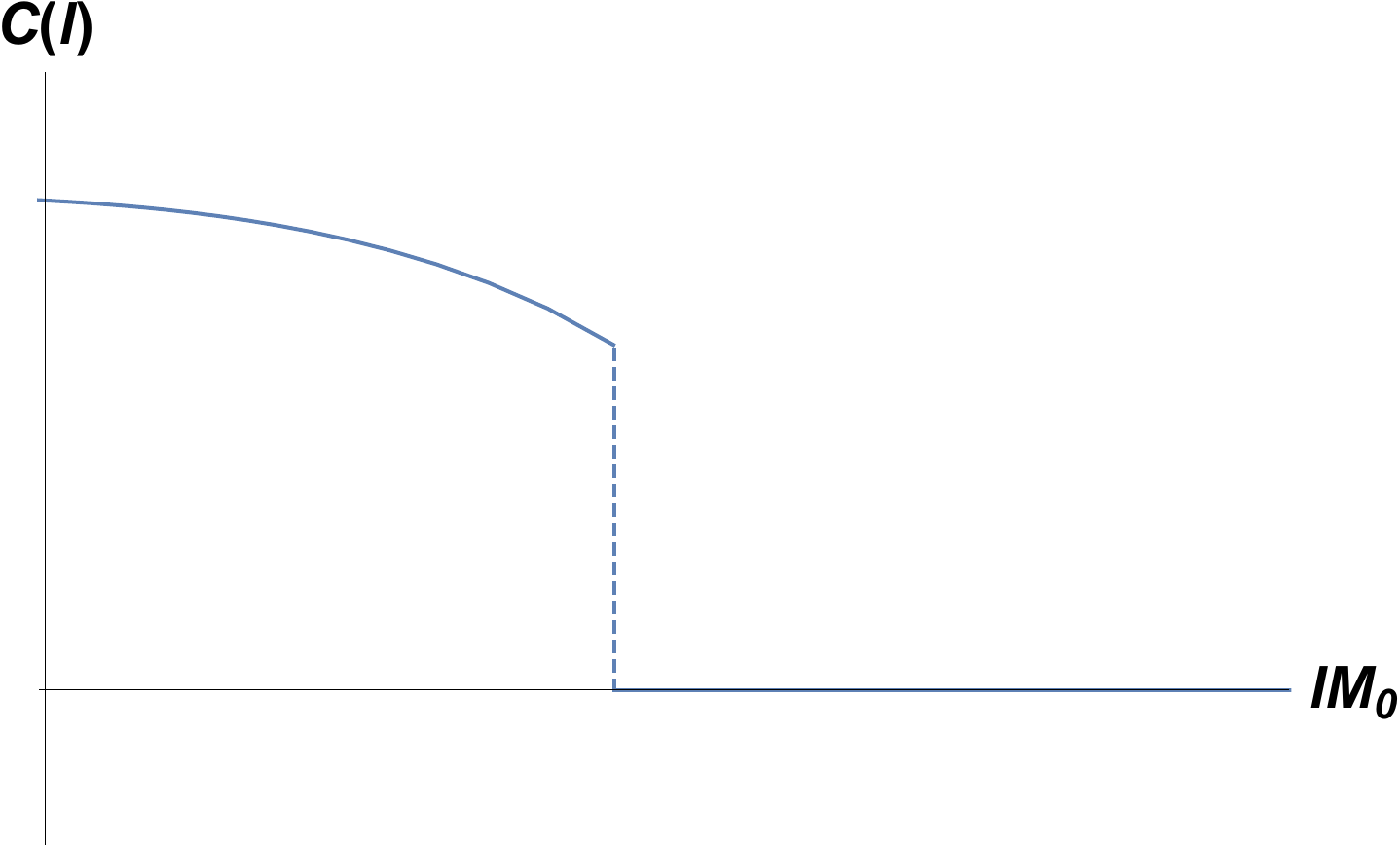}
           \caption{The entropic c function as a function of $lM_0$ at small $a_\phi$. $C(l)$ decreases with increase of $lM_0$. At the critical point $l=l_c$, it jumps and becomes zero. This figure explains the confinement/deconfinement phase transition. Finally, this figure implies that DOF of entangling states are frozen in the confined phase. }
    \label{fig:clike}
    \end{center}
   \begin{center}
  \includegraphics[height=6cm,clip]{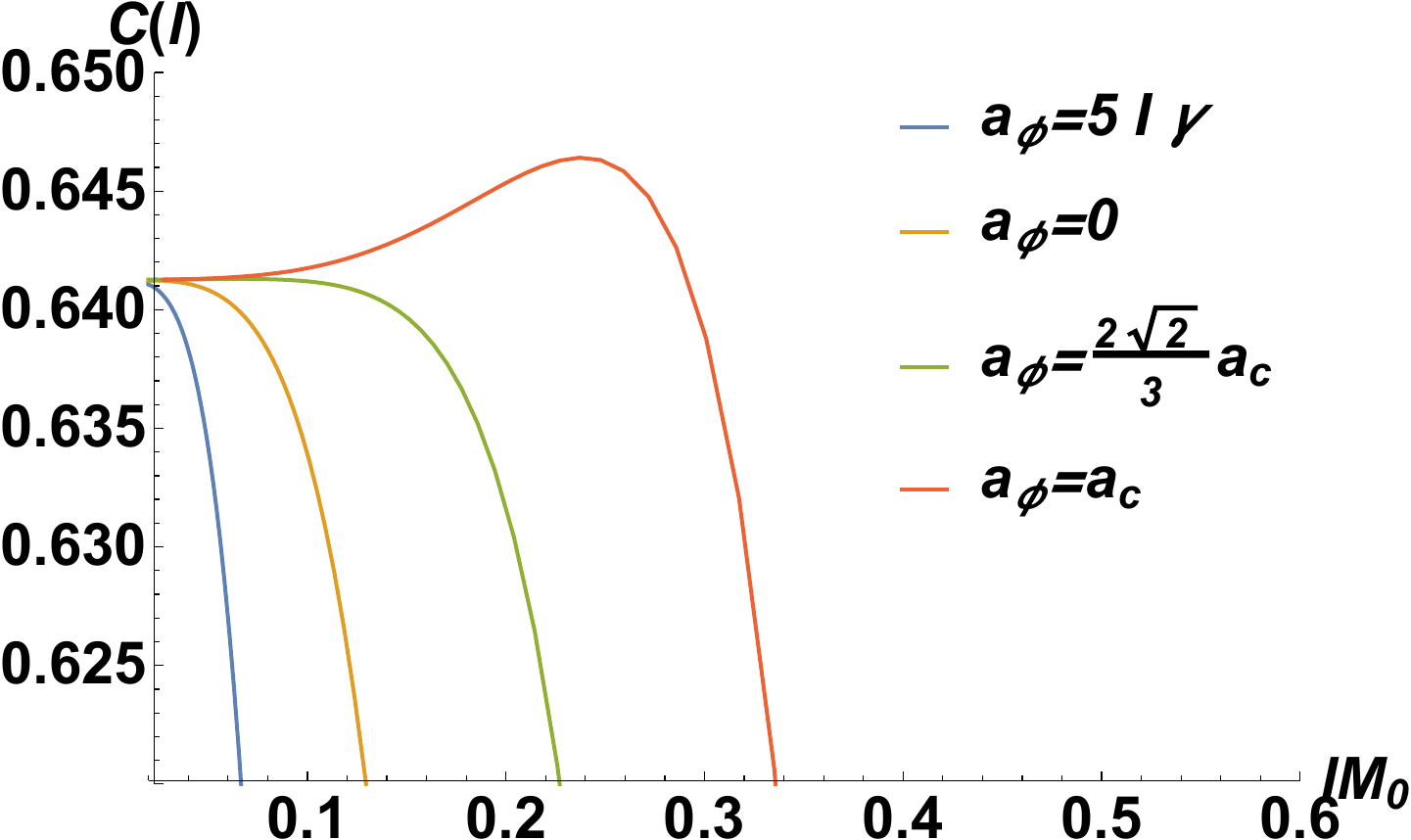}
           \caption{ The entropic c function as a function of $lM_0$ with some fixed $a_\phi$. $C(l)$ can increase until the middle when $a_\phi=a_c$. It decreases after passing the peak. {The green line is a critical line due to the analytical calculation. Lines above the green line increase at first, and line below the green line decreases.}}
    \label{fig:Ecf}
    \end{center}
\end{figure}
The entropic c function shows degrees of freedom at an energy scale $1/l$.
For the pure $AdS$, the c function becomes
\ba
C(l)=C_0\equiv {\dfrac{L^{d-1}2^{d}\pi^{\frac{d+1}{2}}}{\kappa^2}}\Big(\dfrac{\Gamma (\frac{d}{2(d-1)})}{\Gamma (\frac{1}{2(d-1)})}\Big)^{d-1}.
\ea
The entropic c function is plotted as a function of $lM_0$ in Fig. \ref{fig:clike} and \ref{fig:Ecf}. When $a_\phi$ is small or an imaginary number, the entropic c function decreases with increase of $lM_0$. The entropic  c function suddenly becomes 0 at the critical point $l=l_c$. The figure \ref{fig:clike} implies that there are no DOF in the confined phase.
In Fig.  \ref{fig:Ecf}, the entropic c function increases until the middle of the horizontal line. After passing the peak, it decreases.

There is a characteristic length of the system $1/M_0$ which is the radius of the spacial cycle denoted by $S^1$ in the soliton background, the entanglement entropy counts the degrees of freedom (DOF) of the entangled states at the energy scale of the subsystem size $l$ ($E\sim 1/l$) between the subsystem and the complement. Once the subsystem size $l$ chosen here is smaller than the characteristic length ($E \gtrsim M_0$), the entanglement entropy can detect the effective degree of freedom encoding on the $S^1$.  While these DOF can not be detected by entanglement entropy once the subsystem size is much larger than the characteristic length  ($E \ll M_0$), the DOF will be smeared and the entanglement entropy can not identify such kinds of DOF hidden in $S^1$. In this sense, then entanglement c function in the small subsystem can monotonically increase vs the length of the subsystem. However, once the subsystem size is comparable with the characteristic length of the system ($E \sim M_0$), the DOF can not be detectable which leads to the entropy c function monotonically decrease with respect to the size of the subsystem. That is a one interpretation of the non-monotonic behavior of entropy c function with respect to subsystem size. The other interpretation of the non-monotonic behavior of the entropy c function is that the Lorentz symmetry on the boundary of the AdS soliton background has been explicitly broken~\cite{Casini:2006es}. In this sense, the monotonic behavior of the entropy c function can not be protected.

\subsection{HEE in a small region ($d=3$)}
Since we are interested in a small subregion, we expand the action and $l$ in terms of small $z_*$. Near the $AdS$-soliton boundary neglecting the information of the infrared region, $z_*\ll z_+$. The leading order contribution of the surface comes from the $AdS$ boundary and is the zero temperature entanglement entropy in the infinite volume.  Since the background gauge field and the Kaluza-Klein (KK) mass (or finite volume corrections) are small, one can use the perturbation in terms of $z_*$: $M_0z_*\sim M_0l\ll 1$ and $a_\phi l\ll 1$.

We compute small deviations in $d=3$. The size of the interval $l$ is expanded in a power series as follows:
\ba\label{EOM320}
&l=2z_* ( E(-1)- K(-1))+\dfrac{\bar{a}_\phi z_*^4 \left(\frac{1}{2} E(-1) -\frac{1}{2} K(-1)-\frac{\pi  }{8}\right)}{z_+^3} \nonumber \\
&+\dfrac{2 (\bar{a}_\phi -1) z_*^5 (E(-1)-K(-1))}{5 z_+^4}+\dfrac{ \bar{a}_\phi^2z_*^7 \left(\frac{5}{16} E(-1)-\frac{5}{21}  K(-1)-\frac{\pi }{8}\right)}{z_+^6} \nonumber \\
&+\dfrac{ (\bar{a}_\phi-1) \bar{a}_\phi z_*^8  \left(\frac{1}{2}(E(-1)-K(-1))-\frac{1}{64} \pi  \right)}{z_+^7}\dots ,
\ea
where $\bar{a}_\phi =1+\epsilon (\frac{z_+ a_\phi}{\gamma})^2$. $E(k)$ and $K(k)$ are elliptic integrals.
The action is expanded in a power series as follows:
\ba\label{EOM321}
&\dfrac{A}{L_xL^2}=\dfrac{2}{\epsilon}+\dfrac{2 (K(-1)-E(-1))}{z_*} \nonumber \\
&+\dfrac{\bar{a}_\phi z_*^2 \left(\frac{1}{2} E(-1) -\frac{1}{2} K(-1)-\frac{\pi }{4}\right)}{z_+^3}
+\dfrac{z_*^3 \left(\bar{a}_{\phi }-1\right) \left(E(-1)-K(-1)\right)}{z_+^4} \nonumber \\
&+\dfrac{\bar{a}_{\phi }^2z_*^5 \left(\frac{3}{16} E(-1) -\frac{1}{12} K(-1) -\frac{\pi }{8}\right)}{z_+^6}+\dfrac{z_*^6 \left(\bar{a}_{\phi }-1\right) \bar{a}_{\phi } \left(12 E(-1)-12 K(-1)-\pi \right)}{16 z_+^7} \nonumber \\
&-\dfrac{3 \left(\bar{a}_{\phi }-1\right){}^2 z_*^7  \left(E(-1)-K(-1)\right)}{20 z_+^8}+\dfrac{ \bar{a}_{\phi }^3z_*^8 \left(\frac{7}{64} E(-1)+\frac{1}{48} K(-1) -\frac{3 \pi  }{32}\right)}{z_+^9}+\dots
\ea
Note that $z_+=\frac{d}{4\pi M_0}+O(a_\phi^2)$ when $a_\phi\ll M_0$. The above expansion is in terms of the KK mass when $a_\phi =0$. For non-zero $a_\phi$, the above result also shows corrections of  background gauge fields.

HEE is written up to order $l^5$ as follows:
\ba
&{\dfrac{\kappa^2}{2\pi}{S}}=A=\dfrac{2L_xL^2}{\epsilon}-\dfrac{8 \pi ^3L_xL^2}{ \Gamma \left(\frac{1}{4}\right)^4l} \nonumber \\
&+L_xL^2\Big( -\dfrac{4 \Gamma \left(\frac{5}{4}\right)^4 \bar{a}_{\phi }l^2}{\pi ^2 z_+^3}+{\dfrac{3 \Gamma \left(\frac{1}{4}\right)^4 \left(\bar{a}_{\phi }-1\right)l^3}{80 \pi ^3 z_+^4}\Big)}+O(l^5).
\ea
For very small $l$, the increased amount of HEE becomes
\ba
\Delta S= - \dfrac{2L_xL^2 \Gamma (\frac{5}{4})^4\bar{a}_\phi l^2}{{\kappa^2} \pi z_+^3},
\ea
where $\Delta S$ becomes negative when $a_\phi < \frac{2\pi M_0\gamma}{d-1}$. Because quarks can not be isolated in confinement, confinement decreases degrees of freedom. {The $O(l^3)$ contribution coming from the real Wilson line $a_\phi$ is also negative and similar to HEE of the Reissner Nordstr{\"o}m $AdS$ black hole~\cite{Sun:2016til}.}

The energy difference in the dual field theory is defined as
\ba
\Delta E=T_{tt}^{(0)}L_xl=-\dfrac{L_x l L^2  \bar{a}_\phi }{{2\kappa^2} z_+^3},
\ea
where $T_{tt}^{(0)}$ is defined in \eqref{Ttt2} and $\Delta E$ is proportional to the small volume $lL_x=l/M_0$.

The relation like the first law of thermodynamics is as follows:
\ba
\Delta E=T_{ent}\Delta S,
\ea
where entanglement temperature~\cite{Bhattacharya:2012mi,Allahbakhshi:2013rda} is defined as
\ba\label{TEN12}
T_{ent}=\dfrac{\pi}{16  \Gamma (\frac{5}{4})^4 l}.
\ea
Entanglement temperature is inversely proportional to the length $l$. Its coefficient is known to be universal in asymptotic $AdS$ black holes~\cite{Kundu:2016dyk}. However, it is different from $AdS$ solitons  by a factor 2  because the increased amount of HEE $\Delta S$ generally includes $T_{x^ix^i}$ in a striped subregion~\cite{Blanco:2013joa}. Note that the expansion \eqref{EOM320} and \eqref{EOM321} can be computed until higher orders. HEE becomes of $O(\bar{a}_\phi^2l^5/z_+^{6})$ at the next order.

In appendix, we computed entanglement temperature in other dimensions $(d\ge 5)$. Both the increased amount of HEE become negative as in the energy difference when $a_\phi < \frac{2\pi M_0\gamma}{d-1}$. That is, confinement decreases degrees of freedom.

Our results including 3-dimensional ones $(d\ge 3)$ are summarized as follows:
\ba\label{DEN12}
\dfrac{\Delta E}{\Delta S}\equiv T_{ent}=\frac{(d+1) \Gamma \left(\frac{d+1}{2 d-2}\right) \Gamma \left(\frac{3 d-2}{2 d-2}\right)^2}{\sqrt{\pi } d^2 \Gamma \left(\frac{d}{d-1}\right) \Gamma \left(\frac{2 d-1}{2 d-2}\right)^2}\dfrac{1}{l}.
\ea
The above formula shows that the amount of information inside an interval $l$ is proportional to the energy inside the region surrounded by the entangling surface.

We also evaluate the entropic c function for small $l$. The entropic c function is sensitive to the DOF at the energy scale $E\sim 1/l$. Making use of \eqref{ENT41}, it becomes
\ba\label{CTH44}
C(l)=C_0+\dfrac{2T_{tt}^{(0)}l^{d-1}}{T_{ent}}+O(l^d).
\ea
The energy density $T_{tt}^{(0)}$ is negative when $a_\phi < \frac{2\pi M_0\gamma}{d-1}$. Moreover, $C(l)$ decreases with increase of $l$ ($M_0 l\lesssim 1$). On the other hand, the energy density $T_{tt}^{(0)}$ becomes positive when $a_\phi > \frac{2\pi M_0\gamma}{d-1}$: Wilson lines are of the same order as the KK mass  ($a_\phi l/\gamma \sim M_0 l\lesssim 1$). The entropic c function $C(l)$ increases with increase of $l$. {Here it is not necessary that the c theorem $C'(l)\le 0$ has monotonic behavior in a theory with breaking Lorentz symmetry explicitly. Note that results \eqref{DEN12} and \eqref{CTH44} do not apply to a 3 dimensional soliton with Wilson lines.}

\section{Second order correction to HEE}\label{2NDO}
The Fefferman-Graham (FG) expansion is convenient when we consider the asymptotic expansion of the $AdS$ geometric. In this section, we compute the second order correction to HEE with a spherical entangling surface in terms of the FG expansion. A general metric in FG gauge is
\bea ds^2= {L^2\over z^2} \left(dz^2 +  g_{\mu \nu}(z, x^{\mu}) dx^\mu dx^\nu\right),
\eea
where the $AdS$ boundary is located at $z\sim 0$. We approximate that the circle of the $\phi$ direction is large enough ($M_0 L\ll 1$). Therefore, the boundary metric at $z\to 0$ is almost flat. Note that small $M_0$ limit is consistent with assumptions of the asymptotic $AdS$ geometry $M_0L\ll 1$ and $a_\phi L\ll 1$.
The metric is
\bea g_{\mu \nu}(z,x^\mu)=\eta_{\mu \nu}+ \delta g^{(1)}_{\mu \nu}(z,x^\mu)+ \delta g^{(2)}_{\mu \nu}(z,x^\mu). \eea
We assume that the metric is static. The bulk surface stays at a constant time slice. The embedding scalar is $z=z(x^i)$ only.
With the induced metric $h_{ij}=\frac{L^2}{z^2}(g_{ij}+\partial_i z\partial_j z)$, the area is then
\bea A= \int d^{d-1}x \sqrt{h}=\int d^{d-1}x \frac{L^{d-1}}{z^{d-1}} \sqrt{\det g_{i j }}\sqrt{1+g^{ij}\partial_i z \partial_j z}\eea
To evaluate the leading order correction, we use the solution of the zero-th order $z^{(0)}=\sqrt{R^2-\sum_{i=1}^{d-1} (x^i)^2}$.

The first order is \bea& \delta A^{(1)} =\dfrac{\alpha}{2} \int d^{d-1}x \sqrt{\det g_{ij}^{(0)}}\Big(\sqrt{1+g^{(0)ij}\partial_i z^{(0)}\partial_j z^{(0)}}\text{tr}(g^{(0)-1}\delta g^{(1)}) \nonumber \\
&+\dfrac{\delta g^{(1)ij}\partial_i z^{(0)}\partial_j z^{(0)} }{\sqrt{1+g^{(0)ij}\partial_i z^{(0)}\partial_j z^{(0)}}} \Big),
 \eea
where the term linear to $\partial_i z^{(1)}$ vanishes due to the EOM.
{The second order is
\bea \delta A^{(2)} &=&L^{d-1}\int d^{d-1}x \sqrt{ \det g^{(0)}_{ij}} \Big\{ {2 \alpha^2 g^{(0)ij}\partial_i z^{(2)}\partial_j z^{(0)}\over  {{2}}\sqrt{1+ g^{(0)ij} \partial_i z^{(0)} \partial_j z^{(0)}}}+{
 \alpha^2 g^{(0)ij}\partial_i z^{(1)}\partial_j z^{(1)}\over {{2}}\sqrt{1+ g^{(0)ij} \partial_i z^{(0)} \partial_j z^{(0)}}}\nonumber\\&&-
 \Big({\alpha\over 2} {\delta g^{(1)ij}\partial_i z^0  \partial_j z^0}\Big)\Big({2\alpha {{ g^{(0)ij}}}\partial_i  z^{(1)}  \partial_j z^{(0)} \over 2({1+ g^{(0)ij} \partial_i z^{(0)} \partial_j z^{(0)}})^{3\over 2}}\Big)+{\alpha^2\over 2} \Big({{2\delta g^{(1)ij}\partial_i  z^{(1)}  \partial_j z^0 }\over \sqrt{1+ g^{(0)ij} \partial_i z^{(0)} \partial_j z^{(0)}}}\Big)
 \nonumber\\&&+{\alpha^2\over 2} {\delta g^{(2)ij}\partial_i z^{(0)}  \partial_j z^{(0)}\over \sqrt{1+g^{(0)ij}\partial_i z^{(0)} \partial_j z^{(0)}}}-{\alpha^2 \over {{8}} }{ (\delta g^{(1)ij }\partial_i z^{(0)} \partial_j z^{(0)})^2\over (1+ g^{(0)ij }\partial_i z^{(0)} \partial_j z^{(0)})^{3\over 2} }\nonumber\\&&+{\alpha \over 2} \text{tr}(g^{(0)-1} \delta g^{(1)})\Big( {
2 \alpha g^{(0)ij}\partial_i z^{(1)}\partial_j z^{(0)}\over {{2}}\sqrt{1+ g^{(0)ij} \partial_i z^{(0)} \partial_j z^{(0)}}}+
{\alpha\over 2}{ {\delta g^{(1)ij}\partial_i z^0  \partial_j z^0}\over \sqrt{1+ g^{(0)ij} \partial_i z^{(0)} \partial_j z^{(0)}}}\Big)\nonumber\\&&+\Big({\alpha^2 \over 2} \text{tr}(g^{(0)-1} \delta g^{(2)})-{\alpha^2 \over 4}\text{tr}( g^{(0)-1}\delta g^{(1)})^2 {{+}}{\alpha^2\over 8}\text{tr}^2(g^{(0)-1}\delta g^{(1)})\Big) \sqrt{1+ g^{(0)ij} \partial_i z^{(0)} \partial_j z^{(0)}}\Big\}, \nonumber\\\label{(5.8)}\eea}
 where the profile of the minimal surface is corrected due to the change of the bulk metric. Since the remaining computation is lengthy, it is placed in appendix \ref{APPe}.


\section{Discussion}
We analyzed the confinement/deconfinement phase transition and thermodynamic properties of the holographic entanglement entropy in a soliton background with the current. The phase transition occurs due to the competition of two minimal surfaces as analogous to~\cite{Nishioka:2006gr}.  The phase transition happens at the scale $lM_0\sim 1$ (see Fig.~\ref{fig:DA}). We also computed the entropic c function $C(l)$. It probes a phase transition and counts degrees of freedom at an energy scale $E\sim 1/l$. When $E\gtrsim M_0$, HEE can detect the effective degrees of freedom of entangling states inside the circle. When $a_\phi >\frac{2\pi M_0\gamma}{d-1}$, $C(l)$ increases with increase of $l$ and doesn't comply with the c theorem. {When $E\lesssim M_0$, it can not detect degrees of freedom inside the circle and decrease. Note that  the 2-dimensional entropic c function $C_{d=2}(=ldS/dl)$ satisfies  $dC_{d=2}/dl \le 0$ by applying the strong subadditivity to quantum field theories~\cite{Casini:2004bw,Casini:2006es}.  Since the entropic c function $C(l)$ ($d\neq 2$) is a generalized version, however, it does not need to satisfy the analogous condition.}

We derived the relation as in the first law of thermodynamics. The entanglement temperature is defined in \eqref{DEN12} and becomes an inverse function of $l$. We find that both the boundary energy and the increased amount of HEE  become negative in the field theory side when $a_\phi <\frac{2\pi M_0\gamma}{d-1}$. That is, confinement decreases degrees of freedom. On the other hand, $a_\phi$ can increase degrees of freedom and makes both quantities positive.

Finally, the generic formula for the second order correction to the holographic entanglement entropy with contributions from the deformation of the entangling surface has been given. We apply this formula to a spherical entangling surface and reproduce the resulting second order corrections are the same as ones given in \cite{Blanco:2013joa}.

\section*{Acknowledgments}
MF would like to thank B. S. Kim and T. Takayanagi for useful discussions related to this work. MF is supported by the Natural Science Foundation of China. SH would like to appreciate the financial support from Jilin University and Max Planck Partner group. YS would like to thank to the support from China Postdoctoral Science Foundation (No. 2019M653137).

\appendix

\section{The first law in 5 and 6 dimensional QFTs}
In this section, we analyze the deviation from the infinite volume and $d=5$. $l$ and the action are expanded in a power series as follows:
\ba
&l=\dfrac{2 \sqrt{\pi } \Gamma \left(\frac{13}{8}\right)z_*}{5 \Gamma \left(\frac{9}{8}\right)}+\dfrac{\bar{a}_{\phi } z_*^6}{ z_+^5} \left(\dfrac{\sqrt{\pi } \Gamma \left(\frac{13}{8}\right) }{20 \Gamma \left(\frac{9}{8}\right)}-\dfrac{\sqrt{\pi } \Gamma \left(\frac{5}{4}\right)}{12 \Gamma \left(\frac{3}{4}\right)}\right)+\dfrac{}{}\dfrac{4 \sqrt{\pi } \Gamma \left(\frac{13}{8}\right) \left(\bar{a}_{\phi }-1\right) z_*^9}{45 \Gamma \left(\frac{9}{8}\right) z_+^8} \nonumber \\
&+\dfrac{z_*^{11} \bar{a}_{\phi }^2}{z_+^{10}} \left(-\dfrac{\sqrt{\pi } \Gamma \left(\frac{5}{4}\right)}{16  \Gamma \left(\frac{3}{4}\right)}+\dfrac{9 \sqrt{\pi } \Gamma \left(\frac{13}{8}\right)}{320 \Gamma \left(\frac{9}{8}\right)}-\dfrac{7 \sqrt{\pi } \Gamma \left(\frac{15}{8}\right) }{704 \Gamma \left(\frac{11}{8}\right)}\right) \nonumber \\
&+\dfrac{z_*^{14}\bar{a}_{\phi }(\bar{a}_{\phi }-1)}{ z_+^{13}} \left(-\dfrac{\sqrt{\pi } \Gamma \left(\frac{5}{4}\right)}{84 \Gamma \left(\frac{3}{4}\right)}+\dfrac{\sqrt{\pi } \Gamma \left(\frac{13}{8}\right)}{10  \Gamma \left(\frac{9}{8}\right)}\right)\dots,
\ea
and
\ba
&\dfrac{A}{L_x^3L^4}=\dfrac{2}{3\epsilon ^{3}}-\dfrac{2 \sqrt{\pi } \Gamma \left(\frac{13}{8}\right)}{15  \Gamma \left(\frac{9}{8}\right)z_*^3}
+ \left(\dfrac{\sqrt{\pi } \Gamma \left(\frac{13}{8}\right)}{20 \Gamma \left(\frac{9}{8}\right)}-\dfrac{\sqrt{\pi } \Gamma \left(\frac{5}{4}\right)}{4 \Gamma \left(\frac{3}{4}\right)}\right)\dfrac{z_*^2 \bar{a}_{\phi }}{z_+^5} \nonumber \\
&+ \dfrac{\sqrt{\pi } \Gamma \left(\frac{13}{8}\right)}{5 \Gamma \left(\frac{9}{8}\right)}\dfrac{z_*^5 (\bar{a}_{\phi }-1)}{z_+^8} + \left(\dfrac{\sqrt{\pi } \Gamma \left(\frac{13}{8}\right)}{64 \Gamma \left(\frac{9}{8}\right)}-\dfrac{\sqrt{\pi } \Gamma \left(\frac{5}{4}\right)}{16 \Gamma \left(\frac{3}{4}\right)}-\dfrac{\sqrt{\pi } \Gamma \left(\frac{15}{8}\right)}{64 \Gamma \left(\frac{11}{8}\right)}\right)\dfrac{z_*^7 \bar{a}_{\phi }^2}{z_+^{10}}\dots  ,
\ea
The above expansion is able to be used to compute corrections in terms of Kaluza-Klein mass and the background gauge field.

HEE is rewritten up to order $l^5$ as follows:
\ba
&{\dfrac{\kappa^2}{2\pi}}{S}=A=\dfrac{2L_x^3L^4}{3\epsilon ^{3}}-\dfrac{16 \pi ^2 \Gamma \left(\frac{13}{8}\right)^4L_x^3L^4}{1875 \Gamma \left(\frac{9}{8}\right)^4l^3}  \nonumber \\
&+L_x^3L^4\Big(-\dfrac{25  \Gamma \left(\frac{9}{8}\right)^2 \Gamma \left(\frac{5}{4}\right) \bar{a}_{\phi }l^2}{24 \sqrt{\pi }  \Gamma \left(\frac{3}{4}\right) \Gamma \left(\frac{13}{8}\right)^2z_+^5}
+\dfrac{3125 \Gamma \left(\frac{9}{8}\right)^4 (\bar{a}_{\phi }-1) l^5 }{288 \pi ^2  \Gamma \left(\frac{13}{8}\right)^4 z_+^8}\Big)\dots
\ea

The increased amount of HEE becomes
\ba
\Delta S=-\dfrac{25 \sqrt{\pi} \Gamma \left(\frac{9}{8}\right)^2 \Gamma \left(\frac{5}{4}\right) \bar{a}_{\phi }l^2L_x^3L^4 }{12   \Gamma \left(\frac{3}{4}\right) \Gamma \left(\frac{13}{8}\right)^2 {\kappa^2}z_+^5}.
\ea
$\Delta S$ becomes negative when $a_\phi<\frac{2\pi M_0\gamma}{d-1}$. Because quarks can not be isolated in confinement,  confinement decreases degrees of freedom of entangled states.

The increased amount of energy in the dual field theory is defined as
\ba
\Delta E=-\dfrac{l L_x^3 L^4 \bar{a}_\phi }{{2\kappa^2} z_+^5 }.
\ea

Using the relation like the first law $\Delta E=T_{ent}\Delta S$, entanglement temperature is defined as
\ba
T_{ent}=\frac{6 \Gamma \left(\frac{3}{4}\right) \Gamma \left(\frac{13}{8}\right)^2}{25 \sqrt{\pi }  \Gamma \left(\frac{9}{8}\right)^2 \Gamma \left(\frac{5}{4}\right)l}.
\ea

We analyze the deviation in 6 dimensions. The size $l$ and the action are expanded in a power series as follows:
\ba
&l=\dfrac{\sqrt{\pi }  \Gamma \left(\frac{8}{5}\right)z_*}{3 \Gamma \left(\frac{11}{10}\right)}+\left(\dfrac{\sqrt{\pi } \Gamma \left(\frac{8}{5}\right)}{30 \Gamma \left(\frac{11}{10}\right)}-\dfrac{2 \sqrt{\pi } \Gamma \left(\frac{6}{5}\right)}{35 \Gamma \left(\frac{7}{10}\right)}\right)\dfrac{z_*^7 \bar{a}_{\phi }}{ z_+^6} \nonumber \\
&+\dfrac{5 \sqrt{\pi } \Gamma \left(\frac{8}{5}\right) z_*^{11} \left(\bar{a}_{\phi }-1\right)}{66 \Gamma \left(\frac{11}{10}\right)z_+^{10} } \nonumber \\
&+\dfrac{z_*^{13} \bar{a}_{\phi }^2}{ z_+^{12}} \left(-\dfrac{\sqrt{\pi } \Gamma \left(\frac{6}{5}\right) }{25  \Gamma \left(\frac{7}{10}\right)}+\dfrac{11 \sqrt{\pi } \Gamma \left(\frac{8}{5}\right) }{600 \Gamma \left(\frac{11}{10}\right)}-\dfrac{4 \sqrt{\pi } \Gamma \left(\frac{9}{5}\right) }{325 \Gamma \left(\frac{13}{10}\right)}\right)\dots,
\ea
and
\ba
&\dfrac{A}{L_x^4L^5}=\dfrac{1}{2 \epsilon^4}-\dfrac{\sqrt{\pi } \Gamma \left(\frac{8}{5}\right)}{12  \Gamma \left(\frac{11}{10}\right) z_*^4}+\dfrac{z_*^2 \bar{a}_{\phi }}{ z_+^6} \left(\dfrac{\sqrt{\pi } \Gamma \left(\frac{8}{5}\right)}{30 \Gamma \left(\frac{11}{10}\right)}-\dfrac{\sqrt{\pi } \Gamma \left(\frac{6}{5}\right) }{5  \Gamma \left(\frac{7}{10}\right)}\right) \nonumber \\
&+\dfrac{z_*^8\bar{a}_{\phi }^2}{z_+^{12}} \left(-\dfrac{\sqrt{\pi } \Gamma \left(\frac{6}{5}\right) }{25  \Gamma \left(\frac{7}{10}\right)}+\dfrac{\sqrt{\pi } \Gamma \left(\frac{8}{5}\right)}{100 \Gamma \left(\frac{11}{10}\right)}-\dfrac{\sqrt{\pi } \Gamma \left(\frac{9}{5}\right)}{50 \Gamma \left(\frac{13}{10}\right)}\right)+\dfrac{\sqrt{\pi }  \Gamma \left(\frac{8}{5}\right)z_*^6 \left(\bar{a}_{\phi }-1\right)}{6\Gamma \left(\frac{11}{10}\right)  z_+^{10} }\dots ,
\ea
The above expansion can be used to compute corrections in terms of Kaluza-Klein mass and the background gauge field.

HEE is rewritten up to order $l^6$ as follows:
\ba
&{\dfrac{\kappa^2}{2\pi}} S= A=\dfrac{L_x^4L^5}{2 \epsilon^4}-\dfrac{\pi ^{5/2} \Gamma \left(\frac{8}{5}\right)^5L_x^4L^5}{972 \Gamma \left(\frac{11}{10}\right)^5 l^4} \nonumber \\
&+L_x^4L^5\Big(-\dfrac{9 \Gamma \left(\frac{11}{10}\right)^2 \Gamma \left(\frac{6}{5}\right) \bar{a}_{\phi }l^2}{7 \sqrt{\pi } \Gamma \left(\frac{7}{10}\right) \Gamma \left(\frac{8}{5}\right)^2 z_+^6 } + \dfrac{729 \Gamma \left(\frac{11}{10}\right)^5 }{11 \pi ^{5/2} \Gamma \left(\frac{8}{5}\right)^5} \dfrac{(\bar{a}_{\phi }-1) l^6}{z_+^{10}}\Big)\dots
\ea

The increased amount of HEE becomes
\ba
\Delta S=-\dfrac{18 \Gamma \left(\frac{11}{10}\right)^2 \Gamma \left(\frac{6}{5}\right) \bar{a}_{\phi }l^2L_x^4L^5}{7 \sqrt{\pi } \Gamma \left(\frac{7}{10}\right) \Gamma \left(\frac{8}{5}\right)^2{ \kappa^2} z_+^6 }.
\ea

The increased amount of energy in the dual field theory is defined as
\ba
\Delta E=-\dfrac{L_x^4l L^5 \bar{a}_\phi}{{2\kappa^2} z_+^6 }.
\ea

Using the relation like the first law $\Delta E=T_{ent}\Delta S$,
the entanglement temperature is defined as
\ba
T_{ent}=\frac{7 \Gamma \left(\frac{7}{10}\right) \Gamma \left(\frac{8}{5}\right)^2}{36 \sqrt{\pi }  \Gamma \left(\frac{11}{10}\right)^2 \Gamma \left(\frac{6}{5}\right)l}.
\ea

\section{Second Order Correction in Spherical case}\label{APPe}
This appendix is a brief review of second order corrections in~\cite{Blanco:2013joa}.
In section \ref{2NDO}, we have not considered the deformation of profile which is described by $z(x_i)$. We take the deformation into account and we expand
\bea z(x_i )=z_0(x_i) + \alpha z_1(x_i)+ \alpha^2 z_2(x_i)+..., \eea
where $z_0=\sqrt{R^2-\sum_{i=1}^{d-1}x_i^2}$.
Note that since we are only interested in quadratic corrections to the entanglement entropy, $z_2$ will not make contributions since it appears linearly in the area functional. By performing the variation of the action, we will obtain the equation of motion for spherical case.

From formula~(\ref{(5.8)})~we can divide the second order contribution to three category by the power of~$z_{1} $. In the zero-th order of $z_1$,
 \begin{equation}
\begin{aligned}
&A_{2,0}=\int {d^{d-1}x} L^{d-1}z_{0}^d[{\alpha^2 \over 2} \text{tr}(g^{(0)-1} \delta g^{(2)})-{\alpha^2 \over 4}\text{tr}( g^{(0)-1}\delta g^{(1)})^2+{\alpha^2\over 8}\text{tr}^2(g^{(0)-1}\delta g^{(1)})]\dfrac{R}{z_0}
\\&=\int  {d^{d-1}x} L^{d-1}z_{0}^d\{\frac{1}{2}(\eta^{ij}-\frac{x^ix^j}{R^2})(\frac{1}{2}T_{i\alpha}T^{\alpha}_{j}-\frac{\eta_{ij}}{8(d-1)}T_{\alpha\beta}T^{\alpha\beta})
\\&-\frac{1}{4}(\eta^{ij}-\frac{x^ix^j}{R^2})T_{jk}(\eta^{km}-\frac{x^kx^m}{R^2})T_{mi}~+\frac{1}{8}[(\eta^{ij}-\frac{x^ix^j}{R^2})T_{ji}]^2\}\frac{R}{z_{0}}
\\&=\int  {d^{d-1}x} L^{d-1}z_{0}^dR[\frac{1}{4}T_{i\alpha}T^{\alpha i}-\frac{x^ix^jT_{i\alpha}T^{\alpha j}}{4R^2}-\frac{1}{16}T_{\alpha\beta}T^{\alpha\beta}+\frac{r^2}{16R^2(d-1)}T_{\alpha\beta}T^{\alpha\beta}
\\&-\frac{1}{4}T_{im}T^{mi}+\frac{x^ix^jT_{j}^{m}T_{mi}}{2R^2}
-\frac{1}{4R^4}x^ix^jT_{jk}x^kx^mT_{mi}+\frac{T^2}{8}+\frac{1}{8R^4}(x^ix^jT_{ij})^2-\frac{Tx^ix^jT_{ij}}{4R^2}]
\\&=\int  {d^{d-1}x} L^{d-1}z_{0}^dR[-\frac{1}{16}(1-\frac{r^2}{R^2(d-1)})(T^2_{00}+T_{ij}T^{ij})+\frac{T_{i0}T^{i0}}{8}(1+\frac{r^2}{(d-1)R^2})
\\&~+\frac{x^ix^jT_{i\alpha}T^{i\alpha}}{4R^2}+\frac{1}{8}(T^2-T^2_{x}-2TT_{x})],
\end{aligned}
\end{equation}
where we have made use of $\sqrt{1+ g^{(0)ij} \partial_i z^{(0)} \partial_j z^{(0)}}=R/z^{(0)}$ and
\ba
T\equiv T_i{}^i,\quad T_x\equiv T_{ij}\dfrac{x^ix^j}{R^2}.
\ea
The power of $z_{1}$ appears in the second index of $A_{2,n}$. $A_{2,0}$ does not contribute to EOM of $z_1$.
 Next is the power one of $z_1$ as follows:
\begin{equation}
\begin{aligned}
A_{2,1}&=\int {d^{d-1}x\dfrac{L^{d-1}}{z_0^{d-1}}[-\frac{(d-1){R}z_{1}Tr(g_{0}^{-1}g_{1})}{2z_{0}}}+\frac{z_{0}}{R}g^{(1)ij}\partial_{i}z_{1}\partial_{j}z_{1}+\frac{z_{0}}{2R}Tr(g_{0}^{-1}g_{1})g^{ij}_{0}\partial_{i}z_{0}\partial_{j}z_{1}]
\\&=\int  {d^{d-1}x} L^{d-1}[\frac{z_{0}}{R}\partial_{i}z_{1}x^{j}(T^{ij}-\frac{2x^{i}x^kT^{j}_{k}}{R^2}+\frac{x^ix^kT_{kl}x^lx^j}{R^4})+\frac{RTz_{1}}{2z_{0}}-\frac{Rz_{1}T_{ij}x^ix^j }{2z_{0}}]
\\&=\int  {d^{d-1}x} L^{d-1}\frac{R}{2z_{0}}[T(z_{1}-\frac{z^2_{0}}{R^2}x^i\partial_{i}z_{1})+T_{ij}(\frac{2z^2_{0}x^i \partial^{j}z_{1}}{R^2}-\frac{z_{1}x^ix^j}{R^2}-\frac{z^2_{0}x^ix^jx^k\partial_{k}z_{1}}{R^4})].
\end{aligned}
\end{equation}
The power two of $z_1$ becomes
 \begin{equation}
\begin{aligned}
A_{2,2}&=\int  {d^{d-1}x} \frac{L^{d-1}}{z_{0}^{d-1}}[\frac{z_{0}}{2R}g^{(0)ij}\partial_{i}z_{1}\partial_{j}z_{1}+\frac{d(d-1)}{2}(\frac{z_{1}}{z_{0}})^2\frac{R}{z_{0}}+(d-1)\frac{z_{1}}{z_{0}}\frac{x^i\partial_{i}z_{1}}{R}]
\\&=\int  {d^{d-1}x} \frac{L^{d-1}}{z_{0}^{d}}[\frac{d(d-1)z^2_{1}}{2z^2_{0}}+\frac{z^2_{0}(\partial z_{1})^2}{2R^2}-\frac{z^2_{0}(x^i\partial_{i} z_{1})^2}{2R^4}+\frac{(d-1)}{2}\frac{x^i\partial_{i}z^2_{1}}{R^2}].
\end{aligned}
\end{equation}
The EOM of $z_1$ is obtained from the variation of $A_{2,1}+A_{2,2}$.
 \begin{equation}
\begin{aligned}
\frac{\delta(A_{2,1}+A_{2,2})}{\delta z_{1}}=L^{d-1}\int d^{d-1}x\{\frac{R}{2z_{0}}(T-T_{x})+\frac{R}{z^{d}_{0}}(\frac{d(d-1)z_{1}}{z^2_{0}}+\frac{d-1}{R^2}x^{i}\partial_{i}z_{1})\}.
\end{aligned}
\end{equation}
and
 \begin{equation}
\begin{aligned}
\frac{\delta(A_{2,1}+A_{2,2})}{\partial_{k}z_{1}}&=L^{d-1}\int d^{d-1}x\{\frac{R}{2z_{0}}(-\frac{Tz^{2}_{0}x^k}{R^2}+\frac{2T^{k}_{i}z^2_{0}x^i}{R^2}-\frac{z^{2}_{0}T_{ij}x^{i}x^{j}x^{k}}{R^4})      \\&+\frac{R}{z^d_{0}(\frac{z^2_{0}\partial^{k}z_{1}}{R^2}-\frac{z^2_{0}x^{i}\partial_{i}z_{1}x^k}{R^4}+(d-1)\frac{z_{1}x^k}{R^2})}\},
\end{aligned}
\end{equation}
so that the equation of motion becomes
 \begin{equation}
\begin{aligned}
& \frac{2-d}{2R}z_{0}T+\frac{z_{0}(-2-d)T_{x}}{2R}+\frac{(1-d)z_{1}}{R\cdot z^d_{0}}+\frac{(z^2_{0}-2R^2)X^{k}\partial_{k}z_{1}}{R^3 \cdot z^d_{0}}+\frac{\partial_{k}\partial^{k}z_{1}}{z^{d-2}R}
\\&-\frac{2(x^{k}\partial_{k}z_{1}(x^{i}\partial_{i}z_{0}))}{R^3z^{d-1}_{0}}-\frac{x^{i}x^{k}\partial_{i}\partial_{k}z_{1}}{R^3z^{d-2}_{0}}
\\&= \frac{2-d}{2R}z_{0}T+\frac{z_{0}(-2-d)T_{x}}{2R}+\frac{(1-d)z_{1}}{z^{d}_{0}R}-\frac{x^{k}\partial_{k}z_{1}}{R^3z^{d-2}_{0}}+\frac{\partial_{k}\partial^{k}z_{1}}{Rz^{d-2}_{0}}
 -\frac{x^{i}x^{k}\partial_{i}\partial_{k}z_{1}}{R^3\cdot z^{d-2}_{0}}=0,
\end{aligned}
\end{equation}
where {$z_0=\sqrt{R^2-\sum_{i}^{d-1}x_i^2}$}, $T=T_{i}^{i}$ is constant and $T_x=T_{i,j}{x^ix^j\over R^2}$. One can set $T_{i}^{i}=T$ for convenience. We just want to solve the $z_1$.

We consider an ansatz for $z_1$ of the form $z_1=T f_(r)+ T_{i,j}{x^ix^j}f_2(r)$. If we substitute this ansatz, we will have following equation
 \begin{equation}
\begin{aligned}
&\frac{T(d-2)}{r}f'_{1}+Tf''_{1}-\frac{T\cdot r^2}{R^2}f''_{1}+2Tf_{2}+\frac{4T_{x}f'_{2}R^2}{r}+R^2T_{x}f''_{2}
\\&+\frac{(d-2)R^2T_{x}f'_{2}}{r}-2T_{x}f_{2}-4T_{x}rf'_{2}-T_{x}f''_{2}r^2=\frac{z_{0}^d}{2}[(d-2)T+(d+2)T_{x}]
\end{aligned}
\end{equation}
Comparing modes~$T $~and~$T_{x} $~, we have following equation
 \begin{equation}
\begin{aligned}
\frac{d^{2}f_{2}}{du^{2}}(R^{2}-u^2)-\frac{df_{2}}{du}\frac{(d-1)R^2+4u^2}{u}-2f_{2}=\frac{u^{d}(d+2)}{2},
\end{aligned}
\end{equation}
and
 \begin{equation}
\begin{aligned}
\frac{(R^2-u^2)}{R^2}\frac{d^2f_{1}}{du^2}-\frac{(d-1)}{u}\frac{df_{1}}{du}+2f_{2}=\frac{u^d(d-2)}{2}.
\end{aligned}
\end{equation}
We have let~$u=z_{0}$~, this two equation have following solution
 \begin{equation}
\begin{aligned}
f_{1}=-\frac{R^2 z^d}{2(d+1)},~~f_{2}=-\frac{z^d}{2(d+1)}.
\end{aligned}
\end{equation}
The final answer is to coincide with the second order correction  \cite{Blanco:2013joa} to HEE for spherical entanglement surface.

\end{document}